\documentclass[journal=aamick,manuscript=article]{achemso}

\usepackage{adjustbox}
\usepackage{amsmath}
\usepackage{amssymb}
\usepackage{bm}
\usepackage[labelformat=simple,labelsep=period,skip=3pt]{caption}
\usepackage{chemfig}
\usepackage{comment}
\usepackage{dcolumn}
\usepackage{float}
\usepackage{graphicx}
\usepackage[colorlinks, pdfpagelabels, pdfstartview=FitH, bookmarksopen=true, bookmarksnumbered=true, linkcolor=blue, plainpages=false, citecolor=blue, urlcolor=blue]{hyperref}
\usepackage{lineno}
\usepackage{lipsum}
\usepackage{listings}
\usepackage{makecell}
\usepackage[version=3]{mhchem} 
\usepackage[frozencache=true,cachedir=minted-cache]{minted} 
\usepackage{multirow}
\usepackage{minted}
\usepackage{nicefrac}
\usepackage{siunitx}
\usepackage{stmaryrd}
\usepackage{subcaption}
\usepackage{textcomp}
\usepackage[textsize=footnotesize]{todonotes}
\usepackage{url}
\usepackage[british]{babel}

\DeclareMathAlphabet{\mathcal}{OMS}{cmsy}{m}{n}
\DeclareSIUnit{\atom}{atom}
\DeclareSIUnit{\calorie}{cal}
\DeclareSIUnit{\kcal}{\kilo\calorie}
\DeclareSIUnit{\molar}{M}
\usepackage[utf8]{inputenc}
\usepackage[T1]{fontenc}
\usepackage{mathptmx}
\usepackage{etoolbox}

\newcommand{\readded}[1]{\textcolor{black}{#1}}

\author{Shayantan Chaudhuri}
\affiliation[Warwick-Chemistry]{Department of Chemistry, University of Warwick, Coventry, CV4 7AL, United Kingdom}
\alsoaffiliation[Nottingham]{School of Chemistry, University of Nottingham, Nottingham, NG7 2RD, United Kingdom}
\email{shayantan.chaudhuri@nottingham.ac.uk}

\author{Reinhard J. Maurer}
\affiliation[Warwick-Chemistry]{Department of Chemistry, University of Warwick, Coventry, CV4 7AL, United Kingdom}
\alsoaffiliation[Warwick-Physics]{Department of Physics, University of Warwick, Coventry, CV4 7AL, United Kingdom}
\email{r.maurer@warwick.ac.uk}

\title[]{\readded{Challenges in the} Theory and Atomistic Simulation of \readded{Metal }Electrodeposition}


\begin{document}

\begin{tocentry}
\includegraphics{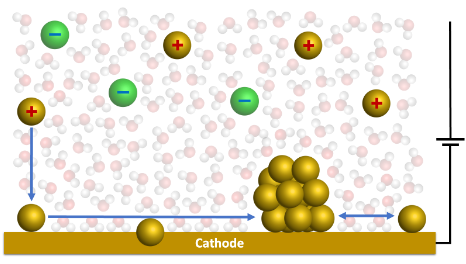}
\end{tocentry}

\begin{abstract}
Electrodeposition is a fundamental process in electrochemistry, and has applications in numerous industries, such as corrosion protection, decorative finishing, energy storage, catalysis, and electronics. While there is a long history of using electrodeposition, its application for controlled nanostructure growth is limited. The establishment of an atomic-scale understanding of the electrodeposition process and dynamics is crucial to enable the controlled fabrication of metal nanoparticles and other nanostructures. Significant advancements in molecular simulation capabilities and the electronic structure theory of electrified solid-liquid interfaces bring theory closer to realistic applications, but a gap remains between realistic applications, theoretical understanding of dynamics, and atomistic simulation. In this review we briefly summarize the current state-of-the-art computational techniques available for the simulation of {electrodeposition} and electrochemical growth on surfaces, and identify the remaining open challenges.
\end{abstract}

\section*{Introduction}

Electrodeposition is the formation of solid structures on the surface of an electrode when an electrochemical potential is applied, and is a viable nanofabrication process alongside more established methods such as nanoimprint lithography~\cite{KoNL07, ChenME06}, pH-driven precipitation~\cite{SharmaCC07, AhnBKCS15}, and directed assembly~\cite{LalanderACSNano10, SharmaAC06, LiL09, LiuNL02}. Metal electrodeposition is a fundamental electrochemical process with many applications such as carbon dioxide reduction catalysis~\cite{MarianoScience17}, water splitting~\cite{KimScience14}, fuel cell applications~\cite{WangNL04}, and materials for energy storage and conversion. In particular, electrodeposition plays a crucial role in battery technologies, where it is used in the controlled deposition of metal electrodes~\cite{KitadaESM25}. Metal electrodeposition is also inherently used in various industrial applications, such as electroplating~\cite{AndricacosIBMJRS98, ChyanJCE08, SchwöbelIOPCSMSE21}, electrowinning~\cite{FinkJES49, IleaHydrometallurgy97, SabaHydrometallurgy00}, and electrocatalysis~\cite{ZhouJACS17, KaleAFM21, LopezCOE21}. Even beyond metals, electrodeposition is commonly used to fabricate non-metallic materials, such as semiconductors~\cite{FulopARMS85}, ceramics~\cite{ZhitomirskyACIS02}, and polymers~\cite{BauerABC23}. \\

Experimental techniques to track, characterize and harness metal electrodeposition have vastly improved over recent years~\cite{LaiCS15, HusseinACSNano18, MoehlLangmuir20}. The complementary use of microscopy approaches, surface spectroscopy methods, and electrochemical analysis provide unprecedented resolution at the nanoscale and, to a more limited extent, resolution in the time domain~\cite{HusseinACSNano18}. \readded{For example, electron microscopy methods have been a popular choice to study metal electrodeposition due to the high resolution they offer~\cite{UstarrozEC10, UstarrozJPCC12, CheriguiJPCC17, UstarrozACSAMI17, HusseinACSNano18}. Scanning transmission electron microscopy has been shown to be capable of dynamically visualizing the early stages of electronucleation for metals, with structural resolution on the atomic scale and time resolution defined by the sequential analysis of short electrodeposition runs~\cite{HusseinACSNano18}. Transmission electron microscopy has also been used to study the electrodeposition of metals at submicroscopic resolution~\cite{UstarrozEC10, UstarrozJPCC12, LeenheerACSNano15, CheriguiJPCC17, UstarrozACSAMI17, WangNL17}. Several studies also report the use of scanning electron microscopy to investigate metal electrodeposition~\cite{PeiNL17, CheriguiJPCC17, LukaczynskaEA19}. While liquid-cell transmission electron microscopy has made lots of progress in monitoring dynamic electrochemical systems~\cite{RossScience15, HodnikACR16}, it has limited resolution due to factors such as electron beam-induced gas bubble formation and electron scattering in the liquid~\cite{HusseinACSNano18}; in contrast, \textit{ex situ} aberration-corrected scanning transmission electron microscopy is not only capable of resolving single atoms but can be used to quantify the number of atoms within a particle~\cite{LiNature08, HusseinACSNano18}. In contrast, scanning probe methods generate images of surfaces using a physical probe that scans the sample~\cite{Khan16}. Both atomic force microscopy~\cite{WangJMCA20} and scanning tunneling microscopy~\cite{LachenwitzerSS97, MöllerPRB97, MöllerZPC99, StrbacPRL99, MorinJES99, LachenwitzerPCCP01, MatsushimaFD16} are capable of analyzing the influence of current on the electrodeposited structure at submicroscopic resolutions. Scanning electrochemical cell microscopy has also been used to study the initial electronucleation stages and mobility of metals~\cite{AaronsonJACS13, LaiCS15, UstarrozACSC18, DaviddiCS21}, and has gained much attention~\cite{EbejerARAC13, BentleyCOE17, BentleyJACS19} due to its ability to routinely operate at submicroscopic scales~\cite{DaviddiCS21, KangLangmuir16, BentleyAC19}. Other techniques such as surface plasmon resonance microscopy~\cite{LaurinavichyuteEA21} and dark-field scattering microscopy~\cite{HillJACS13, HillPCCP13} also exist and have been used to investigate metal electrodeposition. Their performance, however, is restricted by the very small field of view which increases the difficulty in acquiring quantitative data on electronucleation~\cite{LaurinavichyuteEA21, HillJACS13}. Wide-field surface plasmon resonance microscopy, however, removes this constraint and allows for the growth of hundreds of nuclei to be tracked simultaneously~\cite{LaurinavichyuteEA21}.} \\

Simultaneously significant advancements have been made in molecular simulation capabilities and the electronic structure theory of electrified solid-liquid interfaces~\cite{SundararamanJCP17, MelanderJCP19}. Yet a large gap between realistic applications, theoretical understanding of dynamics, and atomistic simulation remains. Challenges that require further understanding include electrode--electrolyte interactions, electronucleation and growth mechanisms, and reaction rates and kinetics. Both theory and experiment face challenges when it comes to bridging this gap and reaching an atomic-level understanding of electrodeposition. Theoretical and computational studies must be able to simulate realistic models capable of replicating experimental conditions, accounting for factors such as the electrochemical potential and surface heterogeneity. On the other hand, model experimental studies should ideally be conducted under well-defined and idealized conditions (e.g. atomically-flat electrode interfaces and well-purified electrolytes) to allow for atomistic simulations and theoretical analyses to be applied~\cite{TrindellCR20}. The synergy between experiment and simulation has the potential to deeply enrich the field, as modeling methods can be refined once information about atomic structure is attained from experiment, while simulations can be used to make predictions that experiments can validate~\cite{TrindellCR20}. The rapid advancement in simulation methodologies has led to a flurry of atomistic simulation activity in the context of electrocatalysis, while much less attention has been given to electrodeposition. Considering its significant industrial relevance and the need for more controlled growth procedures, atomistic simulations and theories that establish a holistic picture of mass transport, reactivity, and growth are urgently needed. It is therefore timely to review modelling methods available for atomistic simulations of electrodeposition. We hope that this review will encourage and guide future efforts in this field and synergize theory and experiment. \\

Existing reviews tend to focus on specific aspects of computational electrochemistry, such as electron transfer processes~\cite{SantosCR22, WarburtonCR22}, modeling methods~\cite{GroßTCC18, Melander21, SundararamanCR22, ZhaoCR22, ZhuCR22, BuiCR22, DattilaCR22, YangWCMS22, YuJMCA23, MelanderEA23, GroßCOE23}, solvation and solid-liquid interfaces~\cite{GroßCR22, RingeCR22, LombardoCR22, YaoCR22}, and the electrochemical double layer~\cite{WuCR22, JeanmairetCR22}. A useful collection of computational electrochemistry reviews is presented in \citet{KoperCR22}. \citet{Gamburg&Zangari11} is a recommended resource that covers the theory and practice of metal electrodeposition, \readded{and} \citet{LinNanoscale24} describes nanoscale phenomena of nucleation and crystal growth in electrodeposition\readded{; however, neither of these resources discuss how computational simulations can contribute towards understanding electrodeposition phenomena}. \readded{E}xisting reviews \readded{either} do not focus on metal electrodeposition in particular, or they \readded{do not} discuss all the relevant elementary processes and aspects that need to be considered \readded{by}  atomistic and continuum simulations. \\

The key challenge in metal electrodeposition is to control the structure, size, and stability of surface-adsorbed nanostructures on an atomistic scale, which in turn define the reactivity and electrochemical properties of the resulting materials. The establishment of an atomic-scale understanding of the electrodeposition process and dynamics is thus crucial to enable the controlled fabrication of metal nanostructures. In this review, we summarize the key concepts and aspects of electrodeposition, as well as the various state-of-the-art computational techniques that are available for the atomistic simulation of electrochemical conditions and electrodeposition processes. Finally, we discuss open questions and challenges in the field.


\section*{Principles of Electrodeposition}

\begin{figure*}[h!]
    \centering
    \includegraphics[width=6in]{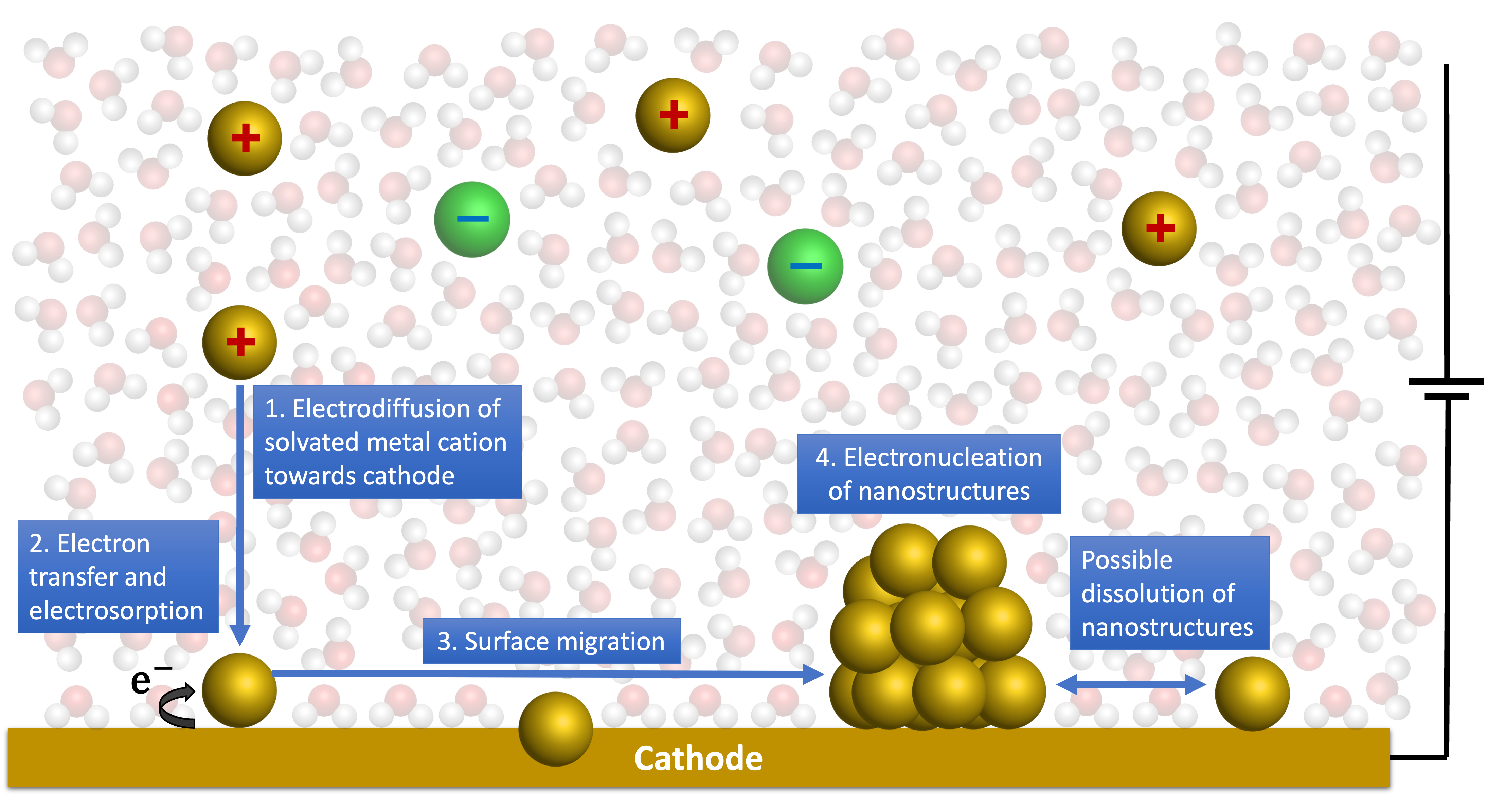}
    \caption{A simplified diagram showing the process of metal electrodeposition onto a cathode surface. First, solvated metal cations electrodiffuse through the solvent (shown as water) towards the cathode (step 1). Once close enough to the cathode, electron transfer will occur that reduces the metal cation, resulting in electrosorption (step 2). The metal atom can then migrate along the cathode surface (step 3), either coming to rest at isolated sites on the cathode surface or coalescing with other metal atoms, resulting in the electronucleation of nanostructures (step 4). The nucleation process is in competition with the dissolution of nanostructures, which can result in further surface migration. Solvated cations and anions are shown as `+' and `$-$', respectively, and hydrogen and oxygen atoms are colored white and red, respectively.}
    \label{fig:Electrodeposition_Summary}
\end{figure*}

The process of metal electrodeposition can be described in four steps (Figure~\ref{fig:Electrodeposition_Summary}): diffusion of metal cations through the solvent due the application of an electric current; electrosorption of metal cations at the cathode surface via electron transfer reactions; migration of metal adatoms along the cathode surface; and the nucleation of larger metal nanostructures on the cathode surface. Electrodeposition can also occur without surface migration, taking place directly at steps or kinks or on adsorbed clusters instead. \\
 
Despite the significant progress in experimental techniques for studying metal electrodeposition, these methods alone cannot fully resolve the underlying physical and chemical processes at play, particularly at the atomic scale. While electron and scanning probe microscopy provide invaluable insights into nucleation dynamics, surface morphology, and growth kinetics, they are often limited by spatial and temporal resolution constraints, as well as the difficulty of directly probing charge transfer and solvent effects in \textit{operando} conditions. Moreover, experimental data typically provide snapshots of the electrodeposition process rather than a complete mechanistic understanding of how ions electrodiffuse, electrosorb, transfer charge, and incorporate into the growing metal phase. To bridge this gap, theoretical modeling and simulation play a crucial role in providing a detailed picture of electrodeposition at multiple length and time scales. Computational methods enable the systematic study of electrodiffusion and electrosorption, electron transfer kinetics, and electronucleation mechanisms, offering predictive power that complements and guides experimental investigations. The following sections outline the key theoretical frameworks used to describe these fundamental processes, highlighting the role of atomistic and continuum-scale simulations in advancing the understanding of electrodeposition.

\subsection*{Electrosorption}
To accurately model metal electrosorption, one must first consider the fundamental differences between adsorption at solid-gas and solid-liquid interfaces. The key difference, in the context of metal electrodeposition, is that once the electrode is negatively polarized, the hydrogen atoms in water molecules and other (partially) positively charged species will be attracted to the electrode surface, and these molecules need to be displaced from the cathode surface before metal cations can adsorb. Adsorption isotherms have been widely used in experiments~\cite{SasakiJES00, LowSCT06, GomesEA06, ThiNamVJC21, DuyenDWT24} to determine parameters such as the maximum adsorption capacity or the relationship between the quantity of particles in a metal deposit and their concentration in an electrolyte. The simplest example is the Langmuir adsorption isotherm~\cite{LangmuirJACS18}, which provides a first approximation of electrosorption behavior, but possesses several inherent assumptions that limit its applicability. It does not take into account the displacement of solvent molecules, assumes the adsorbent surface to be homogeneous, and all adsorption sites are assumed to be energetically equivalent with interactions between adjacent adsorbed species being neglected. Several extensions to the Langmuir adsorption isotherm exist to mitigate these assumptions~\cite{FrumkinZP26, Freundlich26, Bockris00}. While powerful tools, adsorption isotherms require some prior knowledge of the characteristics of the system and do not take into account electrosorption on an atomistic scale. Atomistic simulations can, however, be used to address these limitations by including factors that are excluded within isotherms. \\

The electrosorption of species onto electrode surfaces has typically been described using adsorption theories from gas-surface chemistry, such as $d$-band theory~\cite{Hammer_Nature95}. However, in electrodeposition, d-bands play a minor role, particularly for metals where $d$-bands do not contribute to the density of states (DOS) at the Fermi level. Rather than solely depending on the $d$-band position, the interaction strength depends on the frontier DOS and the alignment of adsorbate orbitals with metal states near the Fermi level~\cite{NørskovPNAS11, huang_spatially_2016}. This concept extends beyond transition metals to $p$-band systems, where partially filled states can similarly influence adsorption. Unlike in gas-surface models, interactions with the electrolyte or solvent molecules significantly affect electrosorption~\cite{LuckyACSNano24}, making a frontier DOS approach more relevant for describing electrodeposition dynamics~\cite{DubouisCS19, ZhangCR22}.

\subsection*{Electron Transfer}
\readded{
During the electrodeposition process, cations undergo electron transfer by which they are reduced to a charge-neutral state.}
Such reactions are considered to be either inner-sphere or outer-sphere. Inner-sphere electron transfer is when the reaction occurs via a strong electronic interaction, such as a covalent bond, and the reactants become connected by a chemical bond. In contrast, outer-sphere electron transfer (OS-ET) occurs between two `unconnected' reactants, and the electron has to move from one reactant to the other through space. This typically occurs at least a solvent layer from the cathode surface~\cite{Taube70}. The mechanism of electron transfer for transition metals can be generally assumed to be outer-sphere as, in solution, transition metal cations can form complex coordination compounds with ligands, such as water molecules, which makes inner-sphere electron transfer less favorable. \\

The charge transfer kinetics at the interface of an electrode are often modeled using the Butler-Volmer formalism~\cite{Butler-Volmer}, which for the case of single electron transfer can be expressed as:

\begin{align}
    k_\mathrm{red} &= k_0\exp\bigg(\frac{-\alpha F\eta}{RT}\bigg) \\
    k_\mathrm{ox} &= k_0\exp\bigg(\frac{(1-\alpha)F\eta}{RT}\bigg)
    \label{eq:Butler-Volmer}
\end{align}
where $k_\mathrm{red}$, $k_\mathrm{ox}$ are the reduction and oxidation reaction rates, respectively, $k_0$ is the rate constant, $\alpha$ is the transfer coefficient that represents the transition state position, $F$ is the Faraday constant, $\eta$ is the applied overpotential, $R$ is the gas constant, and $T$ is the temperature~\cite{SripadJCP20}. The transfer coefficient is generally 0.5 for single-electron processes. In the Butler-Volmer formalism, there is a linear relationship between $\ln{(\nicefrac{k_\mathrm{red}}{k_0})}$ and $\eta$, with a gradient of $\nicefrac{-\alpha F}{RT}$, which is independent of the potential. However, studies have shown that this linearity does not generally hold~\cite{SavéantJECIE75} and that a constant value of $\alpha$ that is independent of the potential cannot universally be assumed~\cite{Bard&Faulkner, LabordaCSR13}. \\

\citet{MarcusJCP56} developed a kinetic model to describe outer-sphere homogeneous electron transfer, in which the rate expressions can be expressed as:
\begin{equation}
    k_\mathrm{red/ox} = k_0\exp\bigg(-\frac{(\Delta G\pm\lambda)^2}{4\lambda RT}\bigg)
    \label{eq:Marcus}
\end{equation}
where $\Delta G$ is the free energy change on reduction and $\lambda$ is the reorganization energy, which requires reorganization of the nuclear configuration of the reactants and solvent to the product state. For multivalent ions, Marcus theory suggests that the simultaneous transfer of multiple electrons is improbable, which implies that the cathodic deposition of multivalent metal cations should occur in a sequence of one-electron steps~\cite{MarcusJCP56, PintoACIE13, PintoCPC14}, with the adsorption reaction being the final one-electron step.  
Marcus theory in combination with molecular dynamics simulations has been used to deduce that small monovalent metal cations, such as Ag$^{+}$, are able to get close to the electrode surface without losing solvation energy as they fit into the water structure well, unlike larger multivalent cations~\cite{PintoACIE13, PhamJPCB18}. As the valency of metal cations increases, the potential of mean force sharply increases on approach to the electrode~\cite{PintoACIE13, PintoCPC14}. When divalent cations approach the electrode surface, they shed their secondary solvation shells, causing their free energy to rise. This means a close approach of divalent (and multivalent) cations to the electrode surface is energetically unfavorable~\cite{PintoCPC14}. \\

One of the important predictions from Marcus theory is that the transfer coefficient $\alpha$ is dependent on the potential. The potential dependence of $\alpha$ can be expressed as:
\begin{equation}
    \alpha = \frac{1}{2} + \frac{F\eta}{4\lambda}
    \label{eq:Potential-Dependent_alpha}
\end{equation}
Potential-dependent transfer coefficients have been observed to agree with experiments~\cite{SavéantJECIE75} and have also been used to explain lithium electrodeposition and stripping data~\cite{BoyleACSEL20}. However, Marcus theory was originally developed for homogeneous electron transfer reactions, whereby the two reactants are present in the same phase. However, for metal electrodes, for example, the charge transfer reaction is heterogeneous~\cite{HenstridgeEA12, LabordaCSR13} and the reaction kinetics will therefore be dependent on electrons from different energy levels within the conduction band of the metal electrode~\cite{MarcusJCP65, hush_adiabatic_1958, hush_electron_1999,ChidseyScience91}. \readded{Based on seminal contributions by Hush, Dogonadze, Kuznetsov, and Gerischer,~\cite{gerischer1960, dogonadze_theory_1965, dogonadze_theory_1970} quantum mechanical rate theories for heterogeneous electron transfer processes have been developed, including what is commonly called the Marcus-Hush-Chidsey model. This model can be expressed with the following rate equation}~\cite{SripadJCP20}:
\begin{equation}
    k_\mathrm{red/ox} = \kappa\int_{-\infty}^{\infty}\exp{\bigg(-\frac{(x-\lambda \mp F\eta)^2}{4\lambda RT}\bigg)\bigg(\frac{1}{1+\exp(\nicefrac{x}{RT})}\bigg)} \mathrm{d}x
    \label{eq:Marcus-Hush-Chidsey}
\end{equation}
where $\kappa$ is a potential-dependent pre-exponential factor that accounts for the attempt frequency in transition state theory and may include other effects such as non-adiabatic corrections, solvent dynamics, and nuclear quantum effects~\cite{!LiuNC21!}. The reader is directed to \citet{HeACIE18} for a more detailed review on the importance of $\kappa$ in electrochemistry. In Equation~(\ref{eq:Marcus-Hush-Chidsey}), $x$ refers to the electronic energy with respect to the Fermi level and the integral captures all electronic contributions to the activation energy weighted by their Fermi-Dirac population, assuming a constant DOS. Marcus-Hush-Chidsey kinetics have been shown to accurately predict Tafel curve data~\cite{BoyleACSEL20, SripadJCP20}. \\

In metal electrodeposition, electron transfer is often coupled with ion transfer, leading to a deviation from purely electronic Marcus-Hush-Chidsey kinetics. Coupled ion-electron transfer theory extends conventional electron transfer models by explicitly considering the role of ion solvation, desolvation, and migration near the electrode surface. The rate of electrodeposition is influenced by both the electronic structure of the electrode and the reorganization of the solvent environment around the electrodepositing metal ion. Such coupling can lead to deviations from classical Butler-Volmer kinetics, especially in multivalent systems where strong solvation effects and complex reaction pathways modify the potential energy landscape. The incorporation of coupled ion-electron transfer theory into Marcus-Hush-Chidsey kinetics provides a more complete picture of metal deposition and dissolution, particularly in energy storage applications where multivalent ions play a role. Mass transport limitations can manifest in experimental voltammetry as deviations from ideal Butler-Volmer or Marcus kinetics. In particular, limiting current densities and diffusion layer effects can lead to non-ideal Tafel slopes. Computational approaches, such as atomistic simulation methods combined with continuum solvation models, can provide insights into how solvation modulates the potential energy landscape. However, connecting these insights, microscopic ET  kinetics, and experimental voltammetric data remains a frontier challenge. \\

Depending on the degree of electronic coupling between the metal nanostructure and the electrode, OS-ET is classified to be either adiabatic or non-adiabatic. Identifying the adiabaticity of OS-ET for different adsorbate-electrode pairs is of fundamental importance in order to optimize the efficiency and mechanism of electron transfer~\cite{LuqueEA13}. In the adiabatic regime, $\kappa$ is independent of the electron tunneling probability between the metal nanostructure and the electrode, and the rate of OS-ET is independent of the electrode material, assuming there exists a sufficiently strong electronic interaction between the adsorbate and the electrode~\cite{HeACIE18, SantosCOE20}. In the non-adiabatic regime, $\kappa$ is proportional to the DOS near the Fermi level. OS-ET reactions on pure metal electrodes are often adiabatic~\cite{LuqueEA13}, while some doubt regarding their adiabaticity on other electrodes remains~\cite{LuqueEA13, NissimCC12, InozemtsevaEA22}. \\

However, \citet{GileadiJEC11} suggested that metal deposition is unique in that charge is carried across the interface by the metal ion, not the electron. This therefore represents a physical situation that is distinctly different from OS-ET. Gileadi proposed a mechanism, based on the assumption that ions migrate across the double layer under the influence of the electrostatic field created by the applied overpotential, using $\alpha$ as the diagnostic criterion. The transfer coefficient quantifies how effectively an applied overpotential lowers the activation barrier for charge transfer, and thus directly influences the nucleation and growth kinetics of metal electrodeposition. Unlike in pure OS-ET reactions, where the solvent dominates the reorganization energy, metal electrodeposition involves strong coupling between the depositing ion and the electrode surface, often with hybridization and partial charge transfer occurring before full incorporation into the growing metal phase. This can lead to a non-trivial dependence of the transfer coefficient on the overpotential and explains observed deviations from the idealized Marcus picture. In particular, the mechanism proposed by \citet{GileadiJEC11} can also explain the observation that heterogeneous rate constants for metal deposition are often higher or comparable to those of OS-ET reactions, despite the fact that bonds are broken during metal deposition to shed the solvation shell, while typically no bonds are broken in OS-ET. Cations at the electrode are rapidly stabilized by interacting with the electrons in the electrode leading to hybridization and partial charge transfer, followed by the slower process of gradual shedding of the solvation shell. During this process, electrodeposition and electrodissolution will remain in competition.

\subsection*{Electronucleation}
Once metal atoms have adsorbed onto an electrode surface, they can start to coalesce to form larger nanostructures in a process known as electronucleation. Individual metal adatoms in partially charged states may migrate along the electrode surface and coalesce to form metastable nanoclusters. Furthermore, larger, crystalline nanostructures can form if smaller structures rearrange and amalgamate together, while closely-spaced nanoclusters can also disassemble and feed atoms into existing nanostructures. In contrast, isolated metal adatoms that are not part of a nanostructure and do not move along the electrode surface might indicate the presence of point defect sites on the substrate surface~\cite{!ChaudhuriJPCC23!}. The potential-driven on-surface dynamics of electronucleation can thus be extremely rich and complex, and atomistic simulations must be able to account for the various thermodynamic and kinetic effects that play a defining part in the size distribution, growth rate, and the rate-determining steps of surface-adsorbed metal nanostructures. \\

When it comes to modeling the process of electronucleation, both classical and atomistic theories exist to describe the formation of stable nuclei~\cite{HusseinACSNano18, SearJPCM07, MilchevCT16, MilchevNanoscale16}. Classical nucleation theory relies on macroscopic physical quantities that are applicable to sufficiently large clusters such that their size can be considered a continuous variable. Common experimental electrochemistry techniques typically provide mostly macroscopic information, from which nanoscopic behavior such as nucleation rates are inferred~\cite{HusseinACSNano18, MilchevJEC76, BudevskiEA00}. Such inferences, however, have been found to be inappropriate to describe the initial stages of nucleation where individual atoms and few-atom clusters are present~\cite{HusseinACSNano18, LaiCS15, UstarrozJACS13, WilliamsonNM03, RadisicNL06, VelmuruganCS12, KimJPCC15}. Classical nucleation theory assumes that the capillarity approximation holds~\cite{SossoCR16, MilchevCT16, MilchevNanoscale16}, which has been shown to break down for small systems~\cite{KiangJAS71, LaaksonenJPCB01, LeeSS73}. Furthermore, clusters are assumed to either grow or shrink via single-atom absorption or emission, respectively, which places kinetic restrictions on the nucleation pathways~\cite{MilchevCT16, MilchevNanoscale16, SossoCR16}. This does not hold in reality as entire clusters can merge or fragment, and these kinetic pathways cannot be ignored. While improvements to classical nucleation theory do exist, such as dynamical nucleation theory~\cite{SchenterPRL99_DNT}, mean-field kinetic nucleation theory~\cite{KalikmanovJCP06_MFKNT}, coupled flux theory~\cite{RussellAM68, PetersJCP11, WeiJAP00, KeltonAM00} and diffuse interface nucleation theory~\cite{GránásyJNCS93, GránásyJNCS95}, these have mostly been applied to describing the condensation of supersaturated vapors into the liquid phase and crystal nucleation studies rather than investigations of metal electronucleation~\cite{SossoCR16, BlowJCP21}. Despite its shortcomings, classical nucleation theory is still a powerful theory and has been shown to be capable of qualitatively capturing nucleation thermodynamics and kinetics for many systems~\cite{SossoCR16}. \\

In contrast, atomistic nucleation theories can be applied to clusters so small that their size can no longer be considered to be continuous, as is the case with first-order phase transitions at high supersaturation levels~\cite{MilchevCT16, MilchevNanoscale16}. Atomistic theories allow for high supersaturation levels to be modeled and have been validated against experimental studies~\cite{MilchevNanoscale16}. Despite the existence of such theories, much remains unclear regarding the initial stages of electronucleation and the role of the atomic-scale structure of the electrode surface. In this regard, explicit atomistic simulations can play an important role in elucidating the initial processes and mechanisms (\textit{vide infra}). \\

The electronucleation of metallic phases is inherently a multiscale problem, spanning atomic-scale processes to macroscopic current transients. A well-established approach to describing electronucleation kinetics involves the analysis of current transients at constant potential, which can provide insights into electronucleation rates and cluster growth dynamics. This formulation, developed through the seminal works of Scharifker~\cite{ScharifkerEA83, ScharifkerJECIE84, MostanyJECIE84, Scharifker14} and Milchev~\cite{MilchevJEC76, MilchevNanoscale16, MilchevCT16}, has provided crucial theoretical frameworks for understanding electronucleation mechanisms. Their models distinguish between instantaneous and progressive electronucleation, describing how nuclei emerge and evolve over time based on the electrochemical conditions. While these approaches have been reviewed previously~\cite{HydeJEC03}, revisiting these key developments is important for contextualizing more recent advancements in atomistic modeling and experimental validation. Linking macroscopic observables, such as current transients, with atomistic descriptions remains a central challenge, highlighting the necessity of multiscale modeling approaches to bridge theoretical and experimental perspectives on electronucleation.

\subsection*{Open Questions Driving the Need for New Simulation Methods}
Despite significant advances in both experimental and computational approaches, a complete and predictive understanding of electrodeposition remains elusive. A major challenge is the current disconnect between atomistic and macroscopic models, which limits the ability to describe the entire deposition process, from the initial stages of adsorption and charge transfer to mesoscopic structure formation and macroscopic film growth. As discussed below, atomistic simulations can provide detailed insights into local interactions, electron transfer~\cite{!LiuNC21!}, and electronucleation at the atomic scale, but are often computationally demanding and struggle to capture long timescale dynamics and large system sizes relevant to real electrodeposition processes. Meanwhile, continuum models~\cite{AlkireJEC03, DrewsAJ04, ZargarnezhadEA17, WeitznernpjCM17, YoonCM18} \readded{(\textit{vide infra})} can simulate extended growth patterns but often rely on parameters that lack direct links to the underlying atomic-scale phenomena. Bridging this scale gap remains a critical open problem. \\

Furthermore, current simulation methods need to incorporate realistic electrochemical environments, such as the influence of solvent dynamics, and the structure of the double layer. Another key challenge is capturing on-surface electronucleation and growth mechanisms with sufficient accuracy. Explicit modeling of the formation of stable growth nuclei and their kinetics remains an area where more computational activity is needed. In addition, the morphology of the electrode can include numerous complexities, such as nanoscale heterogeneities, surface defects, and grain boundaries, which all play a role in localizing charge transfer and influencing electronucleation kinetics. Experimental studies have shown that surface morphology and defect concentration can dramatically alter Tafel slopes, suggesting a direct link between electronucleation events on the atomic scale and macroscopic electrochemical behavior. However, current theoretical models often assume idealized surfaces, limiting their ability to predict realistic deposition kinetics. Developing atomistic models that explicitly account for surface roughness and defects and their impact on charge transfer kinetics remains an important challenge. \\

A key challenge in understanding electrodeposition lies in the disconnect between kinetic theory and experimentally observed kinetics. Kinetic models, such as Butler-Volmer and Marcus-Hush-Chidsey formalisms, describe charge transfer kinetics under idealized assumptions, often neglecting critical factors such as ion solvation, surface heterogeneities, and mass transport effects. Experimentally measured Tafel slopes often deviate from theoretical predictions due to these complexities, particularly under \textit{operando} conditions where solvent interactions and dynamic restructuring of the electrochemical interface play critical roles. \textit{Ab initio} atomistic simulation methods offer a powerful bridge between fundamental kinetic theory and experimental observables and would enable the direct computation of reaction barriers, overpotential-dependent activation energies, and coupled ion-electron transfer mechanisms. By parameterizing kinetic models with \textit{ab initio} data, a more accurate description of reaction pathways and transfer coefficients can be achieved. For example, \textit{ab initio} simulations could quantify how solvation and desolvation contribute to the potential energy landscape, refining the description of coupled ion-electron transfer theory beyond empirical parameterization. Moreover, \textit{ab initio} calculations could provide insights into the potential-dependent nature of the transfer coefficient by explicitly modeling the influence of electronic structure and reorganization energy at the electrode interface. Such an approach would permit a direct comparison with experimental Tafel slopes and voltammetric data, potentially resolving long-standing discrepancies between theoretical predictions and observed electrochemical kinetics. A multiscale approach, via integration of atomistic simulations with experimental measurements, could therefore play a crucial role in advancing understanding of electrodeposition processes and improving the predictive power of kinetic models. \\

Developing better simulation methods will not only improve computational predictions but also provide valuable guidance for experimentalists. Simulations can help interpret complex experimental observations and predict electrodeposition conditions that lead to desirable morphologies or the suppression of defects. By integrating atomistic insights with macroscale modeling, new computational approaches have the potential to unlock more precise control over electrodeposition processes, ultimately advancing technologies in areas such as electrode manufacturing, nanostructured catalysts, and corrosion-resistant coatings. Addressing these open questions will require a concerted effort in developing multiscale modeling frameworks, improved parameterization strategies, and greater synergy between simulations and experiments.

\section*{Simulation of Electrochemical Reaction Conditions}
In this section, we summarize the various state-of-the-art methodologies employed to simulate electrochemical reaction conditions. This includes an accurate account of the electrode potential, the electrolyte, and the electrode surface.

\subsection*{The Electrode Potential}
\begin{figure*}[t]
    \centering
    \includegraphics[width=5in]{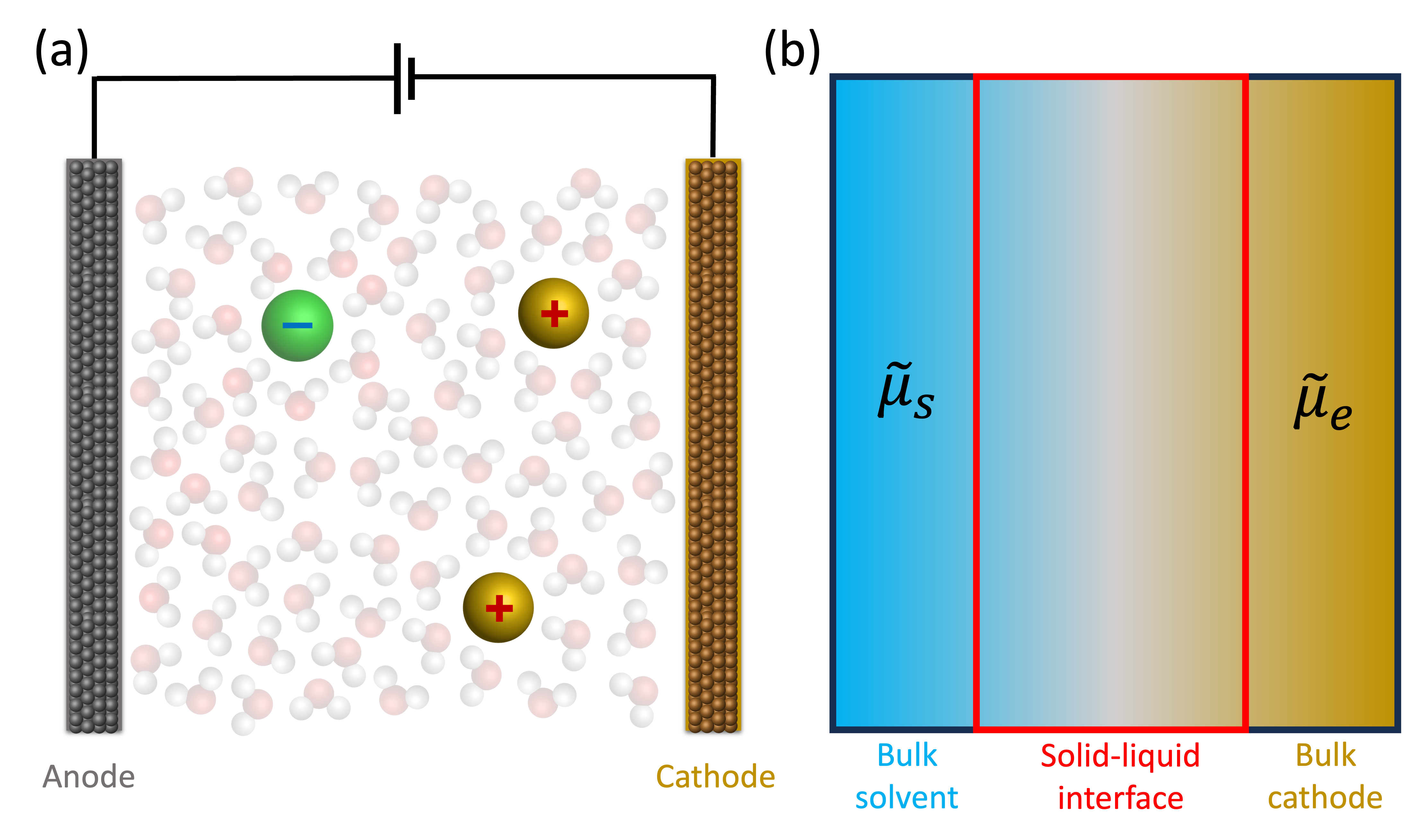}
    \caption{Comparison of (a) full cell and (b) half-cell simulation approaches. Shown in (b) is the interface between the solvent and the electrode in a grand-canonical ensemble, which only requires a half-cell to be simulated. The chemical potentials of the solvent/electrolyte and electrons are fixed to $\Tilde{\mu}_s$ and $\Tilde{\mu}_e$, respectively.}
    \label{fig:Full_versus_Half_Cell}
\end{figure*}

The electrode potential can greatly affect the reaction thermodynamics and kinetics during electrodeposition, and it thus becomes essential to accurately model the potential during atomistic simulations. Electrochemical experiments are typically performed under a constant potential and are referenced against well-defined reference electrodes. However, modeling a constant chemical potential within atomistic simulations is challenging, and various schemes have been developed to control the applied potential~\cite{SakaushiPCCP20, AbidiWCMS21, RingeCR22}. The current schemes to model the applied electrode potential can be roughly split into three categories~\cite{Melander_npj24}: classical force fields~\cite{ScalfiARPC21}, finite-field methods~\cite{ZhangPRB16, DufilsPRL19, DeißenbeckPRL21, JiaJPCL21} which can be used alongside force fields or density functional theory (DFT), and grand-canonical ensembles (GCEs) with an electronic structure method~\cite{SchwarzSSR20, MelanderCOE21} such as DFT~\cite{Hohenberg-Kohn, Kohn-Sham}. The first two methods require the full cell. Force fields do not treat electrons and can only treat electrostatic effects. Grand-canonical treatments, however, only require a half-cell to be modeled with the number of electrons controlled by the applied electrochemical potential. The difference between full- and half-cell approaches is illustrated in Figure~\ref{fig:Full_versus_Half_Cell}. \\

Classical force field and finite-field approaches typically describe the applied electrode potential in terms of the inner potential difference between two electrodes, and do not take into account the electronic structure~\cite{SiepmannJCP95, ReedJCP07, PetersenJPCC12, ZhangPRB16, DufilsPRL19, DwelleJPCC19, CorettiJCP20, TakahashiJCP20, ScalfiARPC21, DeißenbeckPRL21, JiaJPCL21, Melander_npj24}. In both of these approaches, two electrode interfaces need to be simulated in order to build an electrode potential difference within the cell to enforce charge neutrality. However, as the properties of only the working electrode are of interest, the second electrode acts as a passive counter electrode. While the treatment of the additional electrode can be justified for molecular dynamics simulations with force fields, it becomes computationally intractable for electronic structure methods, which do not scale as favorably in terms of their computational cost with the number of atoms in the system~\cite{ZuoJPCA20}. Force fields come in a variety of flavors, including non-reactive and reactive, and often possess a trade-off between accuracy, via highly-detailed parameterizations, and transferability, i.e. the number of materials to which they apply. Force fields are widely used to simulate the energetics and dynamics of large systems by describing their interatomic and intermolecular interactions with a set of parametrized analytical functions. Recently, machine learning (ML) methods have emerged as a successful approach to parameterize force fields by supplying training data from \textbf{ab initio} calculations.~\cite{behler_machine_2021, unke_machine_2021} \\

In contrast, GCE simulations only require one electrode to be modeled and thus enable half-cells to be simulated. This is achieved by fixing the Fermi level of the electrode (which is equal to the chemical potential of electrons, $\Tilde{\mu}_e$), while the chemical potential of the electrolyte, $\Tilde{\mu}_s$, is dependent on the electrolyte solution and its concentration~\cite{MelanderJCP19}. To achieve a well-defined treatment of an electrochemical solid-liquid interface, the electronic structure method (typically DFT) needs to be part of a GCE with a fluctuating number of electrons and ions (at a given temperature), as depicted in Figure~\ref{fig:Full_versus_Half_Cell}, rather than the more common canonical ensemble, where the number of particles is constant but the chemical potentials are allowed to fluctuate~\cite{MelanderJCP19}. Practically, this is most elegantly achieved by fixing the Fermi level and allowing the number of electrons to fluctuate during a calculation~\cite{MelanderJCP19}. While stable algorithms for this method have been developed~\cite{SundararamanJCP17, SmidstrupPRB17}, the fluctuating number of electrons can cause difficulties with convergence in calculations~\cite{MelanderJCP19}. An alternative methodology is to perform calculations with different fixed numbers of electrons and then interpolate to the desired Fermi level~\cite{OtaniPRB06, TaylorPRB06, JinnouchiPRB08, FangJACS10, SkúlasonJPCC10, Letchworth-WeaverPRB12, GoodpasterJPCL16, JinnouchiCOE18}. \\

Changes in the Fermi level directly correspond to changes in $\Tilde{\mu}_e$, which is obtained by changing the charge state of the electrode. This presents a problem for simulations as electrochemical systems are usually partially periodic, but such systems need to be charge neutral. Various methodologies have been proposed to address this, including the introduction of a homogeneous background charge, correction schemes~\cite{TaylorPRB06, DaboPRB08, AndreussiPRB14}, joint DFT~\cite{PetrosyanJPCB05, Letchworth-WeaverPRB12, SundararamanJCP17}, and modified Poisson-Boltzmann implicit solvation models~\cite{Mathew_arXiv16, FisicaroJCP16, RingeJCP17}. \\

\begin{figure}[h!]
    \centering
    \includegraphics[width=2in]{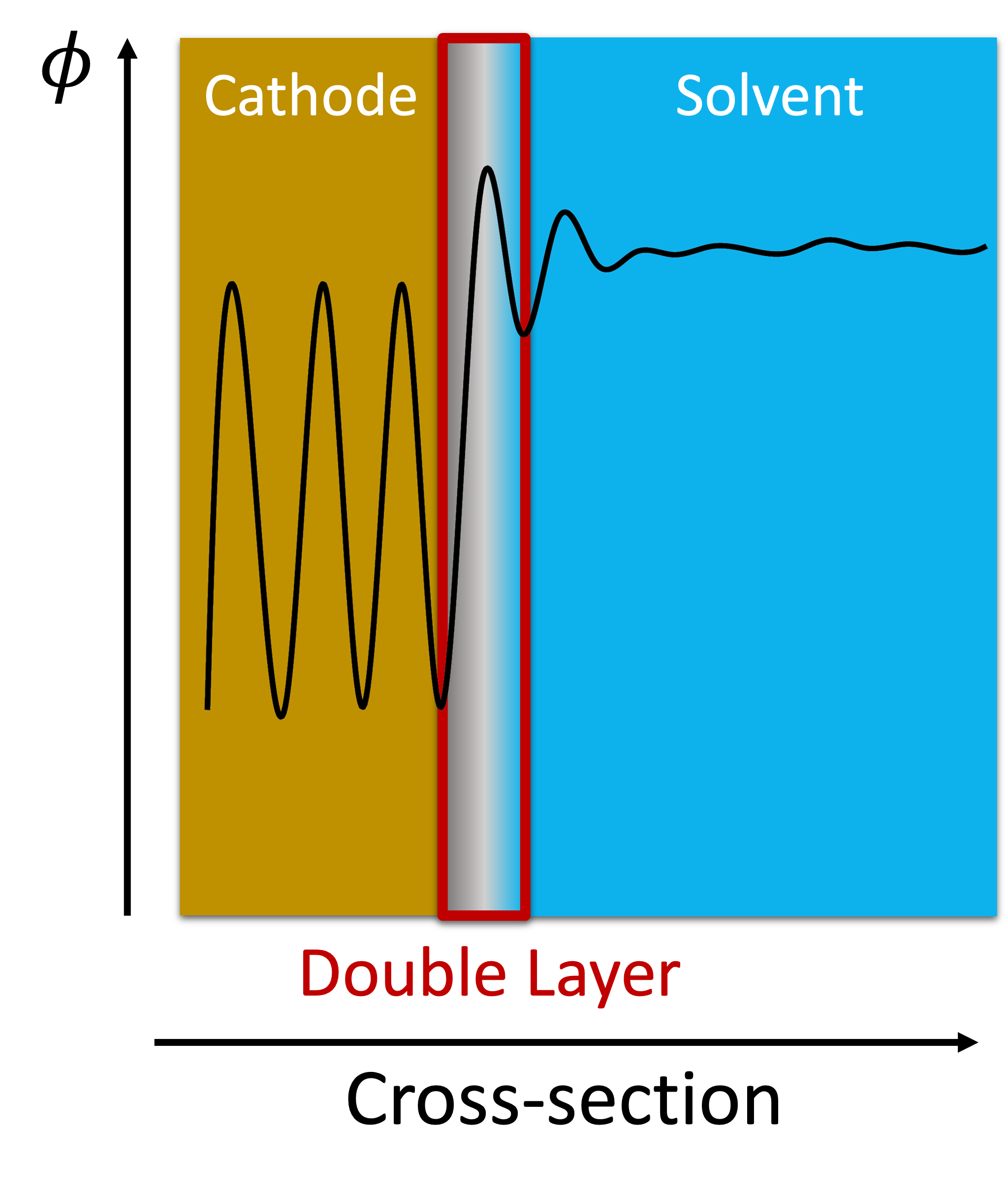}
    \caption{S\readded{implified s}chematic of an example electrostatic potential ($\phi$) profile (black) within the cathode, the double layer, and the bulk solvent.}
    \label{fig:Electrostatic_Potential}
\end{figure}

An important quantity that is often used to analyze how changes in the electrode potential impact the electrochemical system is the electrostatic potential. The redistribution of electrons and ions in the system needed to maintain charge neutrality after changes in the electrode potential is reflected by changes to the electrostatic potential profile, particularly in the double layer region near the electrode. Figure~\ref{fig:Electrostatic_Potential} shows a simplified electrostatic potential profile; at the cathode surface, the electrostatic potential typically exhibits oscillatory behavior due to the periodic arrangement of atoms and the resulting alternating regions of positive and negative charge density. In the bulk solution, the electrostatic potential tends to zero as the overall charge distribution is around neutral due to the lack of nearby charged surfaces and the cancellation of potentials from solvated ions. \readded{However, beyond the metal surface, the electrostatic potential will typically exhibit damped oscillations that reflect the presence of distinct solvation shells around the metal ions. While it can be difficult to assign the exact location of the double layer given the broad continuous nature of the electrostatic potential~\cite{SakongJCP18}, we provide a simplistic visual indication of the double layer region in Figure~\ref{fig:Electrostatic_Potential}.} \\

Fermi level-fixed GC-DFT has been widely used to model electrochemical thermodynamics and kinetics. However, this is not suitable for outer-sphere reactions, semiconductors or two-electrode systems as in these systems, the Fermi level typically lies within the band gap rather than within the conduction or valence bands. To address the shortcomings of Fermi level-fixed GC-DFT, constant inner potential DFT has recently been proposed, which utilizes the local electrode inner potential as the thermodynamic parameter for the electrode potential, rather than the global Fermi level~\cite{Melander_npj24}. Both GC-DFT variants have been shown to provide identical results for metallic electrodes~\cite{Melander_npj24}, but differences can arise for semiconducting metal oxide–water interfaces~\cite{Islas-VargasJCP21}. 

\subsection*{The Electrolyte}
Before solvated cations are electrodeposited onto an electrode surface, they must diffuse through the solvent. Atomistic simulations must therefore ensure they appropriately describe the electrolyte, which includes the solvent itself and charged ions that, upon application of an electrochemical potential, form an electrochemical double-layer that modifies the electrostatic potential above the electrode surface~\cite{MelanderJCP19}. \\

\begin{figure}[h]
    \centering
    \includegraphics[width=3.2in]{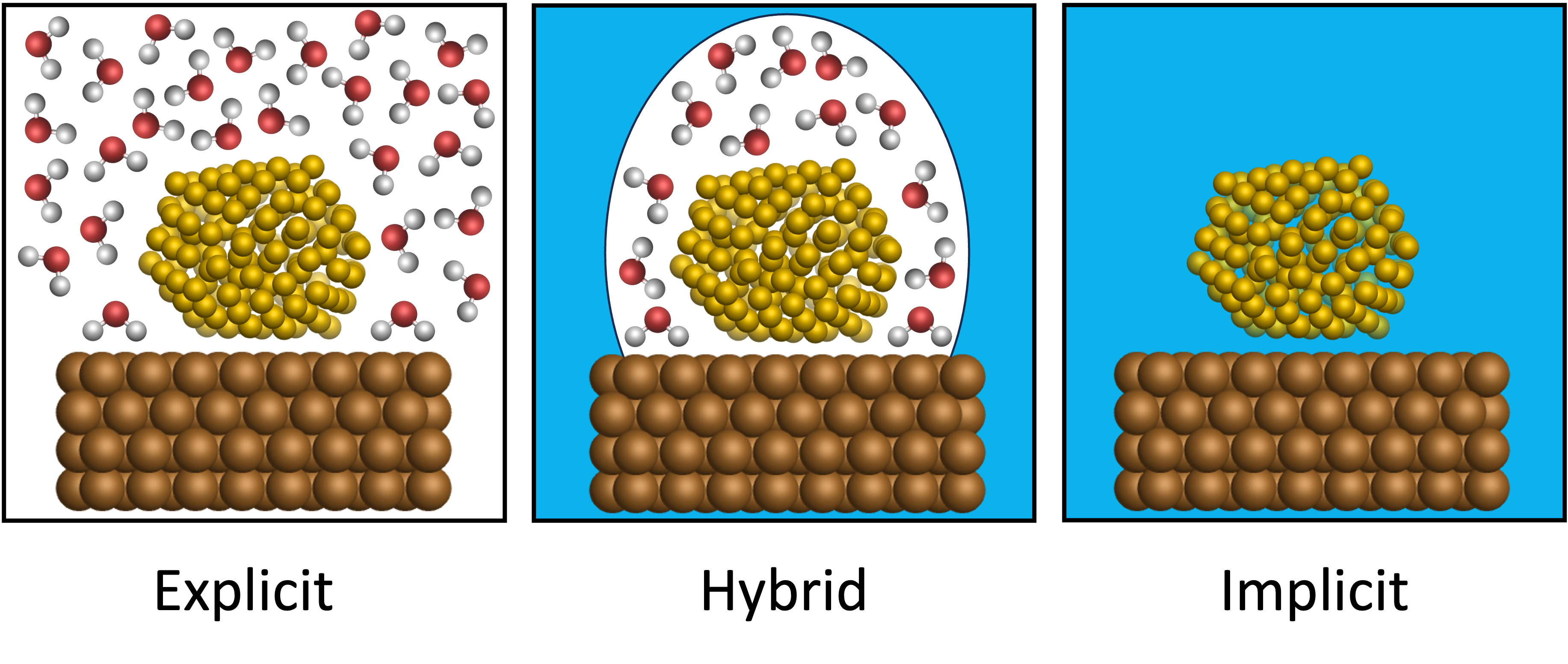}
    \caption{Schematic representations of explicit, hybrid, and implicit solvation environments for a metal nanocluster adsorbed onto an electrode surface within a simulation box. The electrode shown is an atomistic model of a copper (111) surface. Continuum representations of the solvent are shown in blue, while hydrogen, oxygen, adsorbed metal, and copper atoms are colored white, red, gold, and brown, respectively.}
    \label{fig:Solvation_Models}
\end{figure}

Figure~\ref{fig:Solvation_Models} shows the various ways the solvent environment can be described for an electrode surface. Within atomistic simulations, the solvent environment can be described either by using explicit molecules, an implicit (analytical) model, or a hybrid of the two. In the former case, the explicit modeling of solvent molecules is a more `realistic' and physically meaningful description of the system. Specifically in the context of electrodeposition, counter ions such as chloride anions will be present at high concentrations and may need to be explicitly treated. Close to the cathode they will strongly affect the electrostatic interactions, polarization and solvation dynamics during electrodeposition. However, explicit solvation models are typically more computationally expensive than implicit models and in order to model a physically meaningful system, many explicit solvent molecules need to be included within the model, and these can contribute to over 90\% of the atoms within a modeled system~\cite{ZhangJCTC17}. Using quantum mechanical methods to model solute-solute, solvent-solvent, and solute-solvent interactions can therefore quickly become computationally intractable with explicit solvent models comprising many hundreds of atoms. To reduce computational costs, empirical molecular mechanical force fields have become a popular choice to treat interactions within the system. However, care must be taken during the parameterization of such force fields in order to not sacrifice the accuracy that comes with \textit{ab initio} approaches for the sake of computational tractability. In this regard, machine-learned interatomic potentials (MLIPs) offer a lot of promise~\cite{ZhouJPCL23, Zhu_arXiv24}.  \\

In most implicit solvent models, the solvent is treated as an electrostatic continuum with predefined dielectric and interfacial properties~\cite{TomasiCR94, OrozcoCR00, TomasiCR05, ZhangJCTC17}, and the solute is placed inside a cavity within the continuum. In some cases, a dependency on distance from the solute can be included, and it is also possible to introduce a dependence on the rate of a particular process, whereby the response of the solvent varies for fast and slow processes. When using implicit solvent models, a number of choices need to be made, including the shape and size of the cavity that the solute will be placed within. Examples of simple cavity shapes are spherical and ellipsoidal, while more complex ones that can be generated algorithmically such as van der Waals surfaces (based on the van der Waals radii~\cite{BondiJPC64} of atoms), Lee-Richards molecular surfaces~\cite{LeeJML71}, Connolly surfaces~\cite{ConnollyJAC83, ConnollyJMG93}, and surfaces based on charge density isosurfaces~\cite{SinsteinJCTC17} are also possible cavities. \\

One of the advantages of implicit solvent models is that the number of interacting particles and the number of degrees of freedom within a system are significantly reduced. Implicit solvation models are thus typically computationally cheaper than explicit models and are therefore a good practical choice for computationally demanding studies~\cite{ZhangJCTC17, ChenCOSB08, Min_RheePNAS04}. The reduced computational cost also means quantum mechanical methods become more tractable and can be used to treat the solute more accurately than typical molecular mechanical methods. However, the lack of an explicit atomistic description of the solvent can result in numerous interactions, such as hydrogen bonds (both with the solvent and within the solute), being neglected, overstabilized salt bridges, incorrect ion distribution~\cite{LarssonJCTC12}, and unphysical sampling~\cite{ZhangJCTC17, LarssonJCC10, ZhangJCTC15}. Furthermore, the implicit description of the solvent also introduces an artificial boundary between the solute and solvent, and this interface needs to be treated carefully. \\

Typically, the formation of the electrochemical double-layer is modeled using modified Poisson-Boltzmann implicit solvation models, which present the simplest level of fixed-potential GC-DFT for solid-liquid interfaces. Such models come in many varieties, including: linear, non-linear, and with incorporation of ion-size effects. With modified Poisson-Boltzmann solvers, charge neutrality can be maintained via the ionic distribution in the double-layer, though this constraint is not automatically fulfilled in non-linear models~\cite{GuncelerMSMSE13} due to an over-simplification of equations and the presence of the cavity exclusion function~\cite{MelanderJCP19}. This means that when combining modified Poisson-Boltzmann solvers with a quantum mechanical method for charged periodic systems, charge neutrality must be enforced either using Lagrange multipliers~\cite{GuncelerMSMSE13}, a solvated jellium with a constant-charge background tempered by an implicit solvent~\cite{KastlungerJPCC18}, or by treating these systems using `metallic' boundary conditions~\cite{OtaniPRB06}. With a solvated jellium model, a counter charge is dispersed in implicit solvent and a fixed electrochemical potential is iteratively obtained by varying the number of electrons. This works well for metal electrodes, where excess electrons typically localize at the electrode surface and change the work function.~\cite{KastlungerJPCC18}.  \\

Finally, hybrid solvation models that seek to reap the benefits of both explicit and implicit solvation models can also be used to effectively model the electrolyte. In terms of the modeling, hybrid models can be facilitated using embedded cluster models (\textit{vida infra}) and methodologies such as hybrid quantum mechanics/molecular mechanics (QM/MM). Here, the system can be partitioned into three regions, where the central region is treated using QM and comprises the solute and some explicit solvent molecules. This central part is embedded within a second layer and contains more explicit solvent molecules but is treated using MM. Finally, both aforementioned regions are embedded within a third layer, where the solvent is described using implicit models and represents the bulk solvent. This hybrid approach allows the local region of interest to be modeled with the accuracy of QM without subjecting the entire system to the typically higher computational costs that come with QM. However, care must be taken to ensure the various interfaces between the regions are modeled appropriately, and some studies have also observed a dependency on the number of added explicit solvent molecules~\cite{KamerlinCPC09}. 

\subsection*{The Electrode Surface}
\subsubsection*{The Electronic Structure Method}
In order to simulate metal electrodeposition, atomistic models need to include an appropriate description of the electrode surface onto which metals will be deposited. The extended surface model and the choice of quantum mechanical method thus need to correctly account for the rich diversity of interactions that are present at adsorbate-electrode interfaces. Such interactions include hybrid organic-inorganic, long-range van der Waals, and long-range electrostatic interactions of charged species. \\

Kohn-Sham DFT~\cite{Hohenberg-Kohn, Kohn-Sham} is one of the most commonly used electronic structure methods to describe extended surfaces~\cite{HofmannPCCP21} and materials~\cite{MaurerARMS19}. Within DFT, increasingly accurate density-functional approximations are being developed~\cite{LehtolaSX18_Libxc} that can represent the energetics and electronic structure of complex materials. To describe electrified solid-liquid interfaces efficiently, a pragmatic selection of well-tested density-functional approximations that balance computational efficiency and predictive accuracy is required~\cite{MaurerARMS19, HofmannPCCP21}. For example, \textit{a posteriori} long-range dispersion corrections, such as the Grimme~\cite{GrimmeJCC04, GrimmeJCC06, GrimmeJCP10, CaldeweyherJCP17, CaldeweyherPCCP20} and Tkatchenko-Scheffler~\cite{Tkatchenko-Scheffler, RuizPRL12, RuizPRB16, TkatchenkoPRL12, TkatchenkoJCP13, AmbrosettiJCP14, HermannPRL20, !libMBD!} families, to generalized gradient approximations have been shown to provide a reliably accurate representation of adsorption structures and energetics. However, considering the existing limitations in the quantitative experimental characterization of electrochemical systems in general, and the kinetics and dynamics of electrodeposition in particular, the accuracy of existing density-functional approximations and dispersion correction schemes is currently only one of several limiting factor in atomistic electrodeposition simulations. Nevertheless, the choice of quantum mechanical method used remains key for the accurate modeling of processes at electrode surfaces. \\

The intrinsic computational scalability of DFT and even beyond-DFT, such as wavefunction methods and many-body perturbation theory, provides a challenge for systems comprising more than a few hundred atoms, which can often be the case for complex electrode surfaces. Semi-empirical tight-binding methods, such as density functional tight-binding (DFTB)~\cite{ElstnerPRB98_SCC-DFTB}, have grown in popularity as they comprise a good compromise between computational cost and accuracy and have been successfully employed to study the electrodeposition of metals~\cite{LiJPCC18}. However, tight-binding parameterizations are typically developed for a particular subset of elements for a specific purpose, which means they provide low transferability across systems. Furthermore, only a few reliable parameterizations currently exist for metal-organic interfaces~\cite{MoreiraJCTC09_znorg, DolgonosJCTC10_tiorg, FiheyJCC15_auorg, LiJPCC18}, which can be problematic for describing metal electrodeposition on graphite/graphene and boron-doped diamond, or even on multimetallic systems. For these reasons, DFT is still typically more popular than tight-binding methods. However, newer semi-empirical methods, such as xTB, have sought to correct for this drawback by employing a global and element-specific parameterization~\cite{BannwarthWCMS20_xTB, BannwarthJCTC19_GFN2-xTB}, rather than the element pair-specific parameters employed within DFTB. \\

MLIPs are another methodology that can offer high computational efficiency and perform calculations at an \textit{ab initio}-level of accuracy, if the appropriate training data is supplied. Currently, most MLIPs do not sufficiently capture long-range electrostatics and thus will likely not suffice for electrochemical simulations, which require long-range interactions to also be described, though some studies have successfully captured long-range effects within short-ranged MLIPs\cite{MorawietzJCP12, MorawietzJPCA13, YaoCS18_TensorMol-0.1, UnkeJCTC19_PhysNet, KoACR21, KoNC21, !WestermayrDD22!, ZhangJCP22} and further developments on explicit long-range MLIPs are underway.~\cite{UnkeNC21_SpookyNet, kabylda_molecular_2025, caruso_extending_2025} 

\subsubsection*{Structural Model of the Electrode Surface}
For homogeneous electrode surfaces which are atomically flat, periodic boundary conditions can be used as an effective method to model extended surfaces and interfaces~\cite{HofmannPCCP21}. Here, the surface can be represented as a repeated slab, which is trivially defined using a unit cell that is infinitely repeated in three directions. For a surface, it is important to ensure the unit cell possesses a large enough spacing in the $z$-direction to avoid periodic images from interacting with each other~\cite{HofmannPCCP21}. The unit cell should also be large enough to ensure calculations do not suffer from finite size effects~\cite{HofmannPCCP21}; this can be assessed by performing convergence tests on different unit cell sizes. The reader is directed to \citet{HofmannPCCP21} for a more detailed review on the repeated slab approach and practical considerations. \\

However, the surfaces of many electrode materials, particularly semiconducting electrodes, often possess a high degree of heterogeneity and can include structural defects such as point defects, dopants, dislocations, and lower coverages. In fact, defect sites within electrode surfaces have been shown to anchor and stabilize metal nanostructures~\cite{!ChaudhuriJPCC23!, SankarCR20}. This stronger interaction between the nanostructure and the defect site can increase the reactivity exhibited by the supported nanostructure~\cite{DuanPCCP16, EngelPCCP19, CoquetCSR08, ChenCT08, SaqlainPCCP15} in catalytic reactions. Atomistic simulations thus need to be able to account for the heterogeneity of such electrode surfaces. Modeling isolated defects at extended surfaces can be challenging with periodic boundary conditions, as isolated defects within the unit cell acquire a periodicity which can be unphysical. Furthermore, some doped electrode materials typically possess a relatively low dopant concentration e.g. the dopant concentration within boron-doped diamond electrodes is typically around 0.1\%~\cite{MacphersonPCCP15}. To model such an electrode surface using periodic boundary conditions would necessitate a very large unit cell, which can be computationally intractable with higher-level theories. \\

\begin{figure}[h]
    \centering
    \includegraphics[width=3.2in]{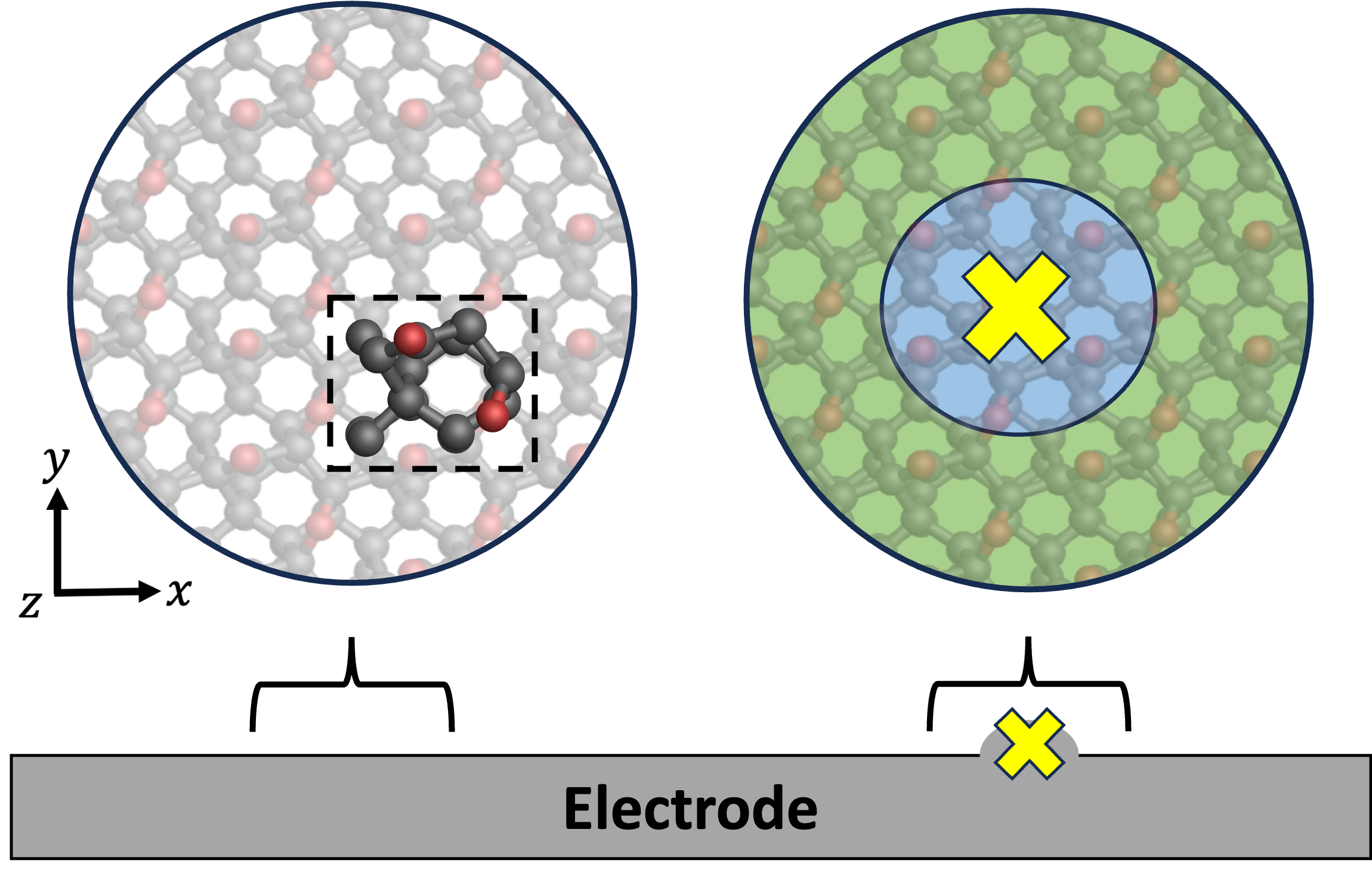}
    \caption{Schematic of how electrode surfaces can be represented within atomistic simulations. For well-defined pristine areas, the electrode can be modeled as a repeated slab via a unit cell to define periodic boundary conditions. For local areas of interest, such as a defect (shown as a yellow X), the surface can be modeled using an embedded cluster approach and partitioned into different regions, shown in blue and green. The electrode surface shown is an atomistic model of an oxygen-terminated polycrystalline boron-doped diamond electrode~\cite{HusseinACSNano18, !ChaudhuriCM22!, !ChaudhuriJPCC23!}.}
    \label{fig:Electrode_Surface}
\end{figure}

The challenges associated with the periodic representation of defects can be overcome by creating truncated cluster models. However, this removes the long-range properties of any bulk material and such calculations can be plagued by spurious finite size effects. Embedded cluster calculations are a viable alternative to periodic slab calculations as they acknowledge the intrinsic locality of surface defect chemistry and permit isolated defects to be modeled whilst breaking translational periodicity. \readded{Figure~\ref{fig:Electrode_Surface} shows how electrode surfaces can be represented within atomistic simulations depending on which region of the surface is of interest.} Embedded cluster models have been treated using a variety of approaches, such as QM/QM~\cite{SauerACR19}, MM/MM and QM/MM, though care must be taken to ensure the embedded region is truncated appropriately, and the interface between the embedded and embedding regions is treated correctly. Such embedded cluster models are also typically computationally cheaper, which makes the application of higher-level theories more straightforward.

\section*{Simulation of Electrodeposition Processes}

Here, we review current atomistic simulation methods and how they capture the elementary steps of electrodeposition described in Figure~\ref{fig:Electrodeposition_Summary}. 

\subsection*{Electrodiffusion \readded{in the Electrolyte and Mass Transport}}
The most common atomistic simulation method to model diffusion is molecular dynamics (MD). MD simulations make use of interaction potentials, and the dynamics of particles (ions, atoms or molecules) are described by numerically integrating Newton's equations of motion, with the forces being computed as the derivatives of the interaction potential. MD simulations that run for long enough should be able to describe the dynamical and structural properties at finite temperatures, as well as the thermodynamic equilibrium. Furthermore, each atomistic trajectory resulting from an MD run allows for the complex mechanisms that drive (electro)chemical processes to be identified. MD simulations, however, cannot easily reach timescales beyond nanoseconds and, without any form of enhanced sampling, cannot describe rare events such as electrodeposition, which can occur over a timescale of seconds. Furthermore, by treating atoms classically, MD alone can only calculate the positions, accelerations and forces of atoms. As protons and electrons cannot be explicitly modeled, key electrochemical processes such as proton conduction cannot be simulated by MD alone. \textit{Ab initio} MD has become popular as it allows for electronic interactions to be included within simulations. Unlike classical MD, \textit{ab initio} MD uses a first-principles method, typically DFT, to calculate forces by solving the time-dependent Schr{\"o}dinger equation. While \textit{ab initio} MD can be more accurate than classical MD, the additional complexity introduced by taking into account the QM interactions results in a much higher computational cost. \\

\readded{To address the limitations of atomistic methods in modeling longer timescales and larger systems, mesoscale simulation techniques such as continuum models~\cite{AlkireJEC03, DrewsAJ04, ZargarnezhadEA17, WeitznernpjCM17, YoonCM18} and kinetic Monte Carlo~\cite{GuoJPCB05, TreeratanaphitakEA14, ChoobarCMS21, ZargarnezhadEA17} play a crucial role in modeling electrochemical reaction conditions and mass transport over larger length and time scales. Continuum models, which typically solve the Poisson-Nernst-Planck equations, provide a macroscopic description of ion transport, charge redistribution, and electrostatic potential gradients within the electrolyte. These models are particularly useful in capturing mass transport effects, including diffusion, migration, and convection, which strongly influence electrodeposition kinetics and morphological evolution. Phase-field methods~\cite{ShibutaSTAM07, ElyJPS14, LiangAPL14, CogswellPRE15, ChenJPS15, LiangEA23, XiongJPS24} introduce order parameters to describe interfacial morphology and dynamic structural changes. Unlike continuum models, phase-field approaches do not track individual atoms but rather model the smooth transition between phases. By incorporating thermodynamic and kinetic driving forces, phase-field approaches enable the study of morphological evolution under different electrochemical conditions, complementing continuum models that focus on ion transport and charge redistribution.} \\

\readded{Mass transport plays a critical role in electrodeposition, particularly in systems where diffusion limitations lead to concentration gradients that influence dendrite formation. For example, the phase-field model developed by \citet{CogswellPRE15} incorporates Marcus kinetics to describe charge transfer in concentrated solutions and allows for simulations of dendritic growth under mass transport constraints. This model, formulated in terms of the grand electrochemical potential, enables the widening of interface regions to reach experimentally relevant length- and timescales whilst enforcing electroneutrality. Notably, the model achieves quantitative agreement with experimental data and predicts that reducing the exchange current, which reflects how readily electrons are transferred between the electrode and ions in solution, can suppress dendrite growth. Screening electrolytes based on their exchange currents could thus serve as a strategy for controlling dendritic instability.}

\subsection*{Electron Transfer Reactions}
Commonly, model Hamiltonians are used to describe interactions between solvated species and the electrode surface and to represent the electronic structure of the system. This is achieved via the inclusion of terms that represent the electronic states of the system and by defining the energy levels of electrons within various species. Other terms that are included within the model Hamiltonian account for Coulombic repulsion (for electron-electron interactions during electron transfer) and hybridization terms (for the overlap between orbitals on the adsorbed metal species and the electrode surface) which can influence the probability of electron transfer occurring. \\

The Newns-Anderson model Hamiltonian~\cite{NewnsPR69, AndersonPR61} is a popular framework that has been used to describe electronic interactions during atomistic simulations of electrodeposition. This framework reduces the complexity of the system while capturing the essential components of electron transfer.   However, there are several limitations of the Newns-Anderson model Hamiltonian, such as the absence of electron correlation and the assumption that the electronic coupling between the electrode and the electrolyte is constant. Solvent effects are also not accounted for, rendering the Newns-Anderson Hamiltonian inappropriate for describing electron transfer processes in highly polarized solvents. Furthermore, the adiabatic approximation is assumed to be valid, where atomic positions are assumed not to instantaneously change with electronic states during electron transfer, which is inappropriate for describing non-adiabatic effects such as coupled electron-proton transfer or dynamically induced electron transfer. \\

Extensions to the Newns-Anderson model Hamiltonian have been developed to explicitly include key microscopic parameters in a single formulation~\cite{SantosCOE20, SchmicklerRJE17, SchmicklerJECIE86, SantosPCCP12, SantosPRB09}. For example, inspired by Marcus theory~\cite{MarcusJCP56}, the Schmickler-Newns-Anderson Hamiltonian~\cite{SchmicklerJECIE86, SchmicklerRJE17} was developed to describe electrochemical electron transfer by incorporating the effect of the solvent. Recently, the Schmickler-Newns-Anderson Hamiltonian has been modified to account for the electrostatic interaction between the electrode and the redox couple~\cite{HuangJPCC18, LamJPCC19, !LiuNC21!} and is expressed as the sum of four terms:
\begin{equation}
    \mathcal{H} = \mathcal{H}_\mathrm{el} + \mathcal{H}_\mathrm{sol} + \mathcal{H}_\mathrm{int} + \mathcal{H}_\phi
    \label{eq:Modified_SNA_Hamiltonian}
\end{equation}
where $\mathcal{H}_\mathrm{el}$ is the electronic contribution and for reactant and electrode surface orbitals ($a$ and $k$, respectively), can be expressed as:
\begin{equation}
    \mathcal{H}_\mathrm{el} = \varepsilon_a(d)\hat{n}_a + \sum_k \varepsilon_k \hat{n}_k + \sum_k \big(V_k(d)c_k^{\dagger}c_a + V_k^*(d)c_a^{\dagger}c_k\big)
    \label{eq:SNA_Hamiltonian_Electronic}
\end{equation}
where $\varepsilon_a$ is the electronic energy of the redox couple, $d$ is the distance between the metal cation and the electrode, $\hat{n}$ is the operator for the occupation number of the redox orbital, and $V_k$ is the interaction parameter that characterizes the strength of electronic interactions~\cite{HuangJPCC18, !LiuNC21!}. The last term in Equation~(\ref{eq:SNA_Hamiltonian_Electronic}) accounts for electron transfer between the metal cation and the electrode surface, with $c^{\dagger}$ and $c$ denoting creation and annihilation operators, respectively~\cite{HuangJPCC18}. As the metal cation approaches the electrode surface, $\varepsilon_a$ will shift towards the Fermi level of the electrode~\cite{HuangJPCC18, SantosPRB09, LangPRB78}. This phenomenon is referred to as Fermi level pinning and occurs due to the stabilization of the cation by the electric field of the cathode and the interaction with the electronic states of the electrode. \\

The solvent contribution for classical nuclei, $\mathcal{H}_\mathrm{sol}$ in Equation~(\ref{eq:Modified_SNA_Hamiltonian}), is given by:
\begin{equation}
    \mathcal{H}_\mathrm{sol} = \dfrac{1}{2}\hbar\omega(p^2 + q^2)
    \label{eq:SNA_Hamiltonian_Solvent}
\end{equation}
where $\hbar$ is the Planck constant, $\omega$ is the solvent frequency, $p$ is the solvent momentum, and $q$ is the solvent reorganization coordinate. The interaction energy, $\mathcal{H}_\mathrm{int}$ in Equation~(\ref{eq:Modified_SNA_Hamiltonian}), between the solvent and the reactant linearly depends on the solvent coordinate and the coupling strength, $g$:
\begin{equation}
    \mathcal{H}_\mathrm{int} = (z-n)\hbar\omega gq
    \label{eq:SNA_Hamiltonian_Interaction}
\end{equation}
Finally, the electrostatic interaction between the redox couple and the electrode surface, $\mathcal{H}_\phi$ in Equation~(\ref{eq:Modified_SNA_Hamiltonian}), can be expressed as:
\begin{equation}
    \mathcal{H}_\phi = (z-n)\phi(d)
    \label{eq:SNA_Hamiltonian_Electrostatic}
\end{equation}
where $z\phi(d)$ represents a repulsive Coulombic interaction between the redox couple center and the electrode and thus possesses an opposite sign to the electrostatic potential obtained from electronic structure calculations, where the electrostatic potential is calculated from the perspective of electrons~\cite{!LiuNC21!}. The right-hand side of Equation~(\ref{eq:SNA_Hamiltonian_Electrostatic}) reflects the change in Coulombic interaction due to reduction and as $\phi(d)$ increases, the stabilization due to reduction increases. \\

The extended Hamiltonian in Equation~(\ref{eq:Modified_SNA_Hamiltonian}) can be parametrized from first-principles using a selection of the techniques discussed above and used to predict both the pre-exponential factor and the free energy barrier in rate equations, and can thus be used to calculate both adiabatic and non-adiabatic ET rates~\cite{SchmicklerRJE17}. As atomistic simulations can be used to obtain all the quantities that enter Equation~(\ref{eq:Modified_SNA_Hamiltonian}), this approach provides physical and chemical understanding of ET kinetics as a function of experimentally accessible parameters. \\

Schmickler-Newns-Anderson-based models have been a powerful theoretical framework for describing OS-ET at interfaces by successfully capturing key aspects of electronic coupling and reorganization energy~\cite{NavrotskayaJCP08, !LiuNC21!, ArguellesJCP24}. \readded{Several studies have combined the model with MD to study electrochemical proton-coupled electron transfer at electrode-solution interfaces~\cite{GrimmingerCPL05, GrimmingerCP07, NavrotskayaJCP08}}. While this approach provides a rigorous way to connect electronic structure theory with electrochemical kinetics, its direct application to electrodeposition remains largely unexplored. The fundamental principles of the model could, in principle, be extended to describe inner-sphere charge transfer and metal adatom stabilization, few studies have pursued this direction. Given the complexity of electrodeposition, where electron transfer is coupled to electrosorption, surface diffusion, and electronucleation processes, extensions of the Schmickler-Newns-Anderson model to explicitly account for these effects would be highly desirable. Future work should explore how modifications of this framework could provide new insights into the electronic structure of intermediates during the electrodeposition process, and how the electrode potential governs electronucleation rates. Developing such extensions could bridge the gap between existing electronic structure models and experimentally observed electrodeposition kinetics.

\subsection*{Electronucleation \readded{and Growth}}
Simulating the amalgamation of surface-adsorbed metal atoms to form metal nanostructures is computationally expensive due to the large number of degrees of freedom and the sheer number of nucleation pathways possible. Some studies investigate the atom-by-atom growth of nanostructures via single-atom absorption, but as mentioned above, this does not hold in reality as entire clusters can merge or fragment. \readded{In this context, it is important to consider the size-dependent surface mobility of nanoclusters, which plays a critical role in determining their likelihood of coalescence or fragmentation. Smaller clusters typically exhibit higher mobility, allowing them to migrate across the electrode surface and contribute to more dynamic growth pathways, whereas larger clusters may become kinetically trapped. These effects can significantly influence both the spatial distribution and the growth kinetics of electrochemical nanostructures. Nevertheless}, analyzing how adsorption energies, cohesive energies, and interatomic distances change as a function of nanostructure size can still provide many valuable insights~\cite{EngelPCCP19, !Thesis!}. \readded{An additional complexity is introduced when attempting to simulate structures and atomic-scale growth under certain electrochemical conditions. \citet{IsaevRSCAdv20} sought to study the growth kinetics of a single hemispherical nanocluster electrodeposited onto an electrode. In particular, they proposed models of formation and diffusion-controlled growth of a new-phase nanocluster for potentiostatic electrodeposition, cyclic voltammetry, and galvanostatic electrodeposition. The nanocluster growth rate during galvanostatic deposition was found to be much lower than under potentiostatic conditions due to the drop in the overpotential that occurs after formation of the nanocluster. However, at larger currents, multiple electronucleation events will occur and increasing the concentration of the depositing cations would cause the number of clusters to decrease and each cluster size to increase.} \\

\readded{F}or a given \readded{cluster size}, metal nanostructures can exhibit numerous metastable geometries. Experimentally, the determination of the geometric ground state of surface-adsorbed nanoclusters is challenging, and simulations are much more promising for providing insights. However, the reliable identification and optimization of surface-adsorbed metal nanostructures is particularly challenging due to the structural complexity and the large number of degrees of freedom, such as the number of possible metastable geometries, adsorption site, and surface coverage~\cite{HofmannPCCP21}. Global structure search approaches such as basin-~\cite{WalesJPCA97, DoyePRL98} and minima-hopping~\cite{GoedeckerJCP04} can be used to identify stable structures. Recently, nested sampling has been extended to calculate coverage-temperature adsorbate phase diagrams by incorporating all relevant configurational contributions to the free energy~\cite{YangPCCP24}. Such stable structure models can be directly compared to \textit{ex-situ} imaging techniques of electrodeposited nanostructures. Combined with sequences of short electrodeposition and \textit{ex situ} imaging, this can provide insights into the structural evolution as a function of macroscopic time.~\cite{HusseinACSNano18}. \\

\readded{\readded{To bridge the gap between atomistic resolution and experimentally relevant timescales, mesoscale methods such as phase-field methods and kinetic Monte Carlo can be applied. Phase-field methods~\cite{ShibutaSTAM07, ElyJPS14, LiangAPL14, CogswellPRE15, ChenJPS15, LiangEA23, XiongJPS24} model the smooth transition between metal and electrolyte phases using order parameters and incorporate both thermodynamic and kinetic driving forces to simulate morphological evolution. For example, they enable simulations of growth, coarsening, and competitive nucleation under varying electrochemical conditions. Kinetic Monte Carlo methods~\cite{GuoJPCB05, TreeratanaphitakEA14, ChoobarCMS21, ZargarnezhadEA17}, on the other hand, provide a statistical framework for modeling the stochastic nature of electronucleation and the impact of thermal fluctuations on electrodeposition. They also enable the time evolution of nucleation and growth, based on transition rates between configurations, to be simulated. This is especially valuable for capturing rare events and long-time processes that are inaccessible to standard MD. The combination of atomistic and mesoscale techniques therefore provides a powerful toolkit for understanding electronucleation and growth mechanisms across length and time scales.}} \\

\readded{\readded{Looking forward, m}esoscale simulation techniques \readded{could} complement atomistic simulations by bridging the gap between detailed quantum mechanical descriptions and experimentally observable macroscopic behavior. While mesoscale models sacrifice some atomic detail, their computational efficiency and ability to capture large-scale mass transport effects make them indispensable for a comprehensive understanding of electrodeposition. By integrating continuum, phase-field, and kinetic Monte Carlo approaches with atomistic insights, robust multiscale frameworks \readded{could} be developed to predict the interplay of ion transport, surface adsorption, and growth. Such a holistic approach \readded{would} not only enhance theoretical understanding but also facilitate the design of more stable, controlled, and efficient electrodeposition processes by optimizing electrolyte composition and deposition parameters.}

\readded{\section*{Conclusions and Outlook}}
Advancements in the theory and atomistic simulation of metal electrodeposition have helped to elucidate many phenomena. Current successes include: accurate modeling of OS-ET reaction kinetics using Marcus-type theories~\cite{SripadJCP20}; continuum simulations that capture mass transport, double-layer formation, and electrostatic potential gradients~\cite{WeitznernpjCM17}; and identification of the role of surface defects and heterogeneities in nucleation through atomistic simulations~\cite{!Thesis!}. However, several open challenges remain that hinder the full realization of holistic models of electrodeposition. For instance, while simulation methods can determine the required overpotential, the exchange current density, and reaction free energies, linking these parameters directly to experimentally measured Tafel slopes, experimentally employed overpotentials and the realistic electrode morphology is still problematic. \\ 

There are notable areas that are still work-in-progress, namely: a unified atomistic simulation framework that captures all aspects of the deposition process, from femtosecond electron transfer events to macroscopic growth kinetics, remains elusive. Such a framework would enable the direct prediction of electronucleation rates, surface morphologies, and growth modes under realistic electrochemical conditions, which currently require empirical input or simplified models. The transition from fundamental charge transfer events to macroscopic deposition patterns remains poorly understood, and a lack of a unified framework would significantly hinder the ability to design materials and processes with tailored electrodeposition characteristics. Secondly, the dynamic nature of the electrochemical interface, including the complex interplay between solvent reorganization, ion pairing, and surface diffusion, is still only partially understood. For example, experimentally observed deviations from classical Tafel behavior in metal electrodeposition, such as unexpectedly high or potential-dependent Tafel slopes, suggest that solvent dynamics and ion rearrangements at the interface play a crucial role in charge transfer kinetics. However, these effects are difficult to disentangle experimentally and remain challenging to model accurately at the atomic scale.  Finally, the coupling of interfacial ion and electron transfer to mass transport in the electrolyte in a predictive, \textit{ab initio} manner is still not fully developed. \\

Long-term challenges include: developing multiscale modeling frameworks that efficiently integrate \textit{ab initio} simulations with mesoscale simulation methods; achieving quantitative agreement between measured Tafel slopes and transfer coefficients from \textit{operando} experiments and computational predictions; and designing robust MLIPs that can accurately capture long-range interactions and electrostatic effects in realistic cathode/electrolyte interfaces. It is evident that MLIPs and other ML surrogate models coupled with electronic structure theory~\cite{WestermayrJCP21} will play a crucial role in developing a unified multiscale modeling description of growth and nucleation at electrified interfaces~\cite{WeitznernpjCM17}. \\

Going forward, specific actions for the computational electrochemistry community should include: strengthening collaborations with experimental groups to validate and refine simulation models through direct comparison with high-resolution \textit{operando} data; developing standardized benchmarks for electrodeposition simulations, focusing on key observables such as overpotential, exchange current densities, electronucleation rates, and morphology evolution; advancing algorithms to bridge the time and length-scale gap, particularly through hybrid multiscale methods that integrate \textit{ab initio} insights with continuum modeling; and focusing on the parameterization of coupled ion-electron transfer theories using \textit{ab initio} data to better reproduce experimental voltammetry and Tafel slopes. \\

In summary, while significant progress has been made, a disconnect remains between detailed atomistic simulations and macroscopic experimental observables. Bridging this gap requires a concerted effort to integrate quantum mechanical insights with multiscale modeling approaches, thereby enabling more accurate and predictive simulations of electrodeposition processes. Crucially, to reduce the complexity gap and facilitate experiment-theory synergy, this will, at least initially, require experimentation on heavily simplified systems and idealized model systems. A necessary detour that will provide significant rewards in terms of improved models and understanding. Enhanced collaboration between theorists and experimentalists is essential to refine models, validate simulation outputs, improve the interpretation of experiments, and ultimately achieve a unified description of metal electrodeposition. \\


Both theory and experiment will be needed to answer the open questions in the field. In terms of electronucleation, this includes consideration of non-classical growth pathways such as surface migration, aggregation and coalescence of small nanoclusters~\cite{LaiCS15, KimJPCC15, HodnikACR16}; formation of metastable clusters into crystalline nanoparticles~\cite{HusseinACSNano18}; and nucleation and dissolution events that occur before stable nuclei form~\cite{HarnimanNC17, SiepmannJCP95}. Other open questions~\cite{Scharifker14, MilchevNanoscale16} include why the measured number of nuclei is higher than the calculated number of active sites and why single atoms are so stable. With the emergence of improved simulation methods and experiments with high spatial and temporal resolution, there is promise for these open questions to be answered in the near future.

\begin{acknowledgement}
This work, in part, builds on S.C.’s doctoral thesis~\cite{!Thesis!}. This work was funded by the EPSRC Centre for Doctoral Training in Diamond Science and Technology [EP/L015315/1], the Research Development Fund of the University of Warwick, Wellcome Leap as part of the Quantum for Bio Program, the UKRI Future Leaders Fellowship programme [MR/S016023/1 and MR/X023109/1], and a UKRI Frontier research grant [EP/X014088/1].
\end{acknowledgement}



\providecommand{\noopsort}[1]{}\providecommand{\singleletter}[1]{#1}%
\providecommand{\latin}[1]{#1}
\makeatletter
\providecommand{\doi}
  {\begingroup\let\do\@makeother\dospecials
  \catcode`\{=1 \catcode`\}=2 \doi@aux}
\providecommand{\doi@aux}[1]{\endgroup\texttt{#1}}
\makeatother
\providecommand*\mcitethebibliography{\thebibliography}
\csname @ifundefined\endcsname{endmcitethebibliography}  {\let\endmcitethebibliography\endthebibliography}{}

\end{document}